\documentclass{article}

\usepackage[utf8]{inputenc} 
\usepackage[T1]{fontenc}    
\usepackage{hyperref}       
\usepackage{url}            
\usepackage{booktabs}       
\usepackage{amsfonts}       
\usepackage{nicefrac}       
\usepackage{microtype}      
\usepackage{subfigure}
\usepackage{graphicx}
\usepackage{enumerate}
\usepackage{makecell}

\title{Approximate LTL model checking}

\author{
  Weijun ZHU*, Jianwei WANG and Yongwen FAN \\
  School of Information Engineering, Zhengzhou University\\
  Zhengzhou City, 450001 China \\
  \texttt{* E-mail: zhuweijun@zzu.edu.cn} \\
}

\begin{document}

\maketitle

\begin{abstract}
  Linear Temporal Logic (LTL) Model Checking (MC) has been applied to many fields. However, the state explosion problem and the exponentially computational complexity restrict the further applications of LTL model checking. A lot of approaches have been presented to address these problems. And they work well. However, the essential issue has not been solved due to the limitation of inherent complexity of the problem. As a result, the running time of the existing LTL model checking algorithms will be inacceptable if a LTL formula is too long. To this end, this study tries to seek an acceptable approximate solution for LTL model checking by introducing the Machine Learning (ML) technique, and a method for predicting results of LTL model checking via the Boosted Tree (BT) algorithm is proposed in this paper. First, for a number of Kripke structures and LTL formulas, a data set A containing their model checking results is obtained, using the existing LTL model checking algorithm. Second, the LTL model checking problem can be induced to a binary classification problem of machine learning. In other words, some records in A form a training set for the BT algorithm. On the basis of it, a ML model M is obtained to predict the results of LTL model checking. As a result, an approximate LTL model checking technique occurs. The experiments show that the new method has the predictive accuracy of 98.0\%, and its average efficiency is 9.4 million times higher than that of the representative model checking method, if the length of each of LTL formulas equals to 500. These results indicate that the new method can quickly and accurately determine results of LTL model checking for a given Kripke structure and a given long LTL formula since the new method avoid the famous state explosion problem.
\end{abstract}

\section{Introduction}

Model checking was presented by Turing Award winner Prof. Clarke et al [1].  And it is a key technique which can verify automatically whether a computing system satisfies a given property. Up to now, model checking has been applied to many fields, such as CPU design [2], security protocols [3] and malware detection [4], and this technique has been used by some leading IT companies, including INTEL and IBM [5].

The basic principle of model checking can be depicted as follows: (1) a finite automaton or a Kripke structure is employed to construct a systematic model, while a formula of a temporal logic is employed to describe a property which should be satisfied by this system; (2) a model checking algorithm decides whether the automaton or the Kripke structure satisfies the formula or not; (3) the result of model checking will be “true”, if the automaton or the Kripke structure satisfies the formula; (4) otherwise, the result of model checking will be “false”. In model checking, linear temporal logic [6], which was introduced to computer science by Turing Award winner Prof. Pnueli, and computational tree logic (CTL) [7][8], which was proposed by Turing Award winner Prof. Clarke, are the two popular temporal logics. And these two logics have been used widely in international IT industry.

The state explosion problem is always one of the important bottlenecks of LTL model checking. To address this problem, many methods including symbolic, partial order reduction, equivalence, compositional reasoning, abstract and symmetry et al [1], have been proposed to reduce the huge state space, which is caused by the model checking algorithms. As a result, these methods work well. In a special case, \(10^{120}\) states were verified automatically by a symbolic model checker [9]. However, the huge state space still restricts the further applications of model checking. The game of Go is a famous example. The huge state space will prevent any Go method from exhausted search for all strategies for the two players, if the model checking technique is employed.

Unfortunately, the state explosion problem inherently originates from the gene of model checking, LTL model checking in particular. Thus, no solution within the framework of hard computing exists. It is well known that soft computing have the following properties: uncertainties, inaccuracies, incomplete true value, low-cost and robust. Is there any better solution in soft computing? This is the open issue. Motivated by it, we will conduct a study in this paper.

\section{Preliminary }

\subsection{LTL model checking \& NuSMV/NuXMV}
The basic principle of LTL model checking can be depicted as follows [1]: (1) the not form of a LTL formula is converted to a Kripke structure; (2) the above Kripke structure intersects another Kripke structure expressing a systematic model; (3) the algorithm will decide that the systematic Kripke structure satisfies the LTL formula, if the intersection set is empty; (4) otherwise, the systematic Kripke structure does not satisfy the LTL formula.

NuSMV is developed by Carnegie Mellon University, University of Genova, University of Trento and FBK-IRST [10]. It is a free tool for symbolic CTL model checking and symbolic LTL model checking. See Ref.[10] for more details on NuSMV.

In addition, NuXMV extends NuSMV. And NuXMV features a strong verification engine based on state-of-the-art SAT-based algorithms.

\subsection{BT algorithm and Graph Lab}
A core goal of Machine Learning (ML) is to classify data. There are many ways applied for classification, such as support vector machines, random forests, decision trees, BT and deep learning. As a class of popular ML algorithms, BT has the following advantages: good effect, insensitive to input and low computational complexity. Thus, BT has been applied to many fields, such as text segmentation [11], face detection [12], hand pose recognition [13], multi-view, multi-pose object detection [14] and emotion recognition [15], etc. In this paper, we use a kind of BT algorithm called Gradient Boosted Regression Trees (GBRT) [16] to conduct our studies.

As one of the most effective ML algorithms, the GBRT algorithm has the strong generalization ability. GBRT will generate multiple decision trees, and the results of all the trees are accumulated to form the final answer. The core of this algorithm is that each tree learns from the residuals of all previous trees. GBRT can be employed to deal with not only some regression problems but also some binary classification problems. If the latter problem is dealt with, a threshold will be set. A logical 1 will be gotten if the value of regression computation is greater than the threshold. Otherwise, a logical 0 will be gotten.

The advantage of GBRT lies in [17]: (1) Strong ability in handling mixed type of data; (2) Strong predictive ability; (3) Strong robustness against outliers. The disadvantage of GBRT is that [17] parallel processing cannot be performed.

Graph Lab is an open source ML package [18], which was developed by Carnegie Mellon University. This tool integrates a variety of ML algorithms including GBRT, which greatly simplifies the training process of the model, and facilitates users’ operations and implementation of specific ML algorithms.
\section{The principle of the new method}
We consider the following specific problem introduced in section 1. How to determine whether systematic model K satisfies a LTL formula f or not, giving a pair of K and f using a ML approach.

\begin{figure}
\centering
\subfigure[ for a given pair of Kripke structure K and a LTL formula f, determine whether K satisfies f or not]{
\begin{minipage}[b]{0.5\textwidth}
\includegraphics[width=\textwidth]{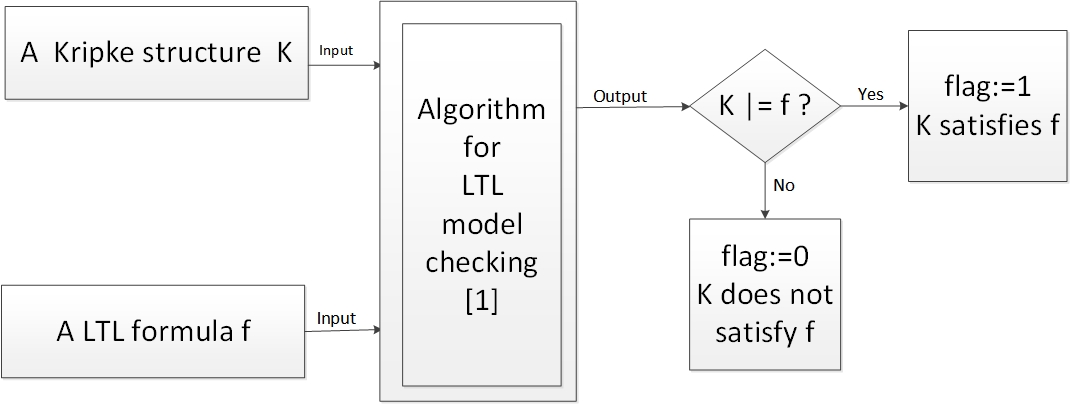} \\
\label{fig1-1}
\end{minipage}
}

\subfigure[the model M can predict the model checking results for \(m_2-m_1\) pairs of K and f, since M is obtained by training \(m_1\) groups of K, f and their model checking result r ]
{
\begin{minipage}[b]{0.5\textwidth}
\includegraphics[width=\textwidth]{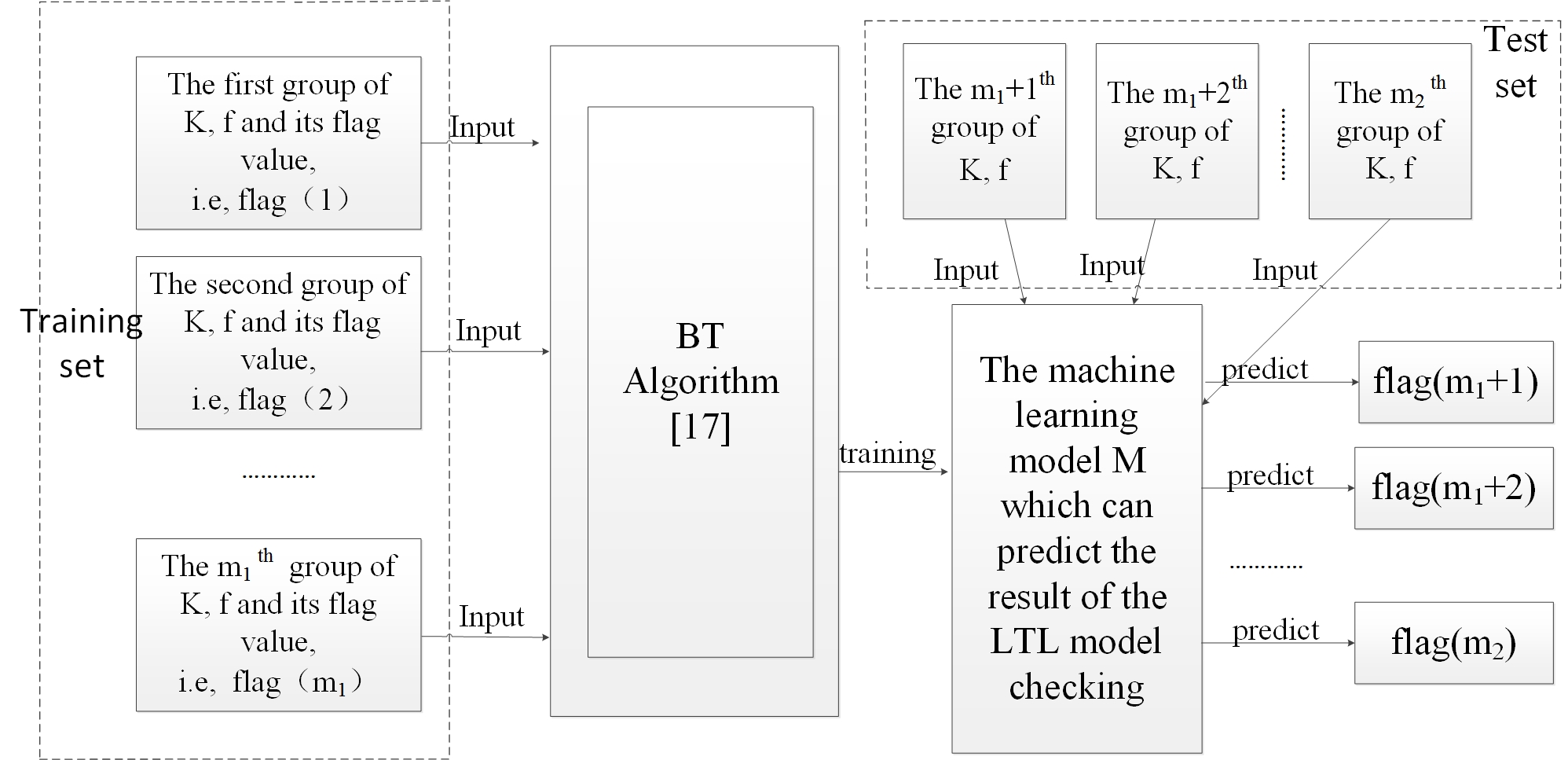} \\
\label{fig1-2}
\end{minipage}
}

\caption{given one pair of model and formula, the new method can determine/predict whether this model satisfies this formula or not}
\label{fig1}
\end{figure}

The principle of our method is shown in Figure~\ref{fig1}. The core is to train with a number of records containing information on systematic models, LTL formulas and their model checking results, using BT algorithm. Thus, a ML model called M which has a predictive ability is obtained. And then, a given pair of K and f is input into the model M. As a result, the output of M is the predicted result indicating whether K satisfies f or not.

The steps of the process can be described as follows.
(1) As shown in Figure~\ref{fig1-1}, one can run a LTL model checking algorithm and obtain a result of model checking for a given pair of K and f. On the basis of it, he or she can perform a binary classification. The result of the classification will be 1, if the result of model checking is true. Otherwise, the result of the classification will be 0.
(2) Step (1) is repeated $m_1$ times, and a training set containing $m_1$ records is gotten. See the left part of Figure~\ref{fig1-2}.
(3) Train using the BT algorithm with the training set obtained in step (2), and the ML model M is obtained. See the middle part of Figure~\ref{fig1-2}.
(4) Another pair of K and f which is required to predict their model checking result, is input to the trained model M. Whether K satisfies f or not? It can be predicted by M.

\section{Simulated experiments}
\subsection{The objective of the experiments}
We will explore the ability and the efficiency of the new method based on ML. Specifically, can the new method improve the efficiency significantly under the premise that the new method can approach the popular LTL model checking one in terms of power?
\subsection{The simulation platform}
(1) CPU: Intel(R) Core(TM) i7-4790 CPU @3.60GHz.(2) RAM: 8.0G RAM.(3) OS: Windows 10.(4) NuSMV and NuXMV: for performing LTL model checking.(5) Graph Lab: for implementing the BT algorithm.

\begin{figure}
\centering
\includegraphics[width=0.4\textwidth]{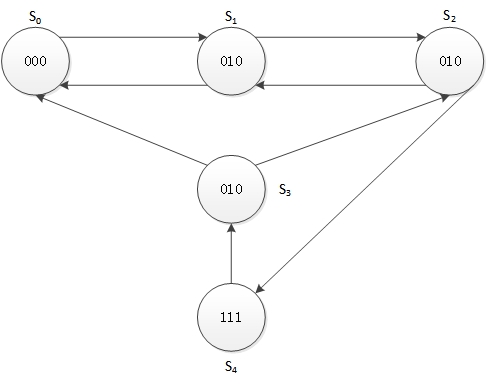}
\caption{an example on Kripke structure \(K=K_0\) }
\label{fig2}
\end{figure}

\begin{figure}
\centering
\includegraphics[width=\textwidth]{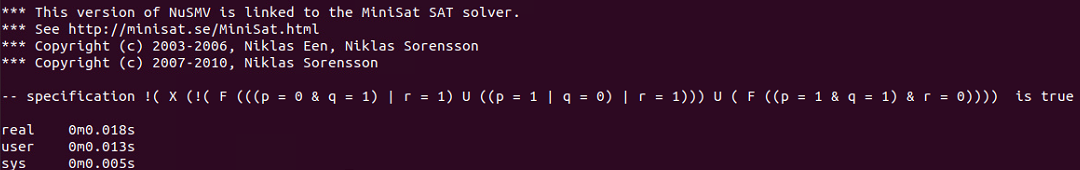}
\caption{an example on MC, \(K_0\) satisfies the LTL formula \(f_1\), where length of \(f_1\) is 25}
\label{fig3}
\end{figure}

\begin{figure}
\centering
\includegraphics[width=\textwidth]{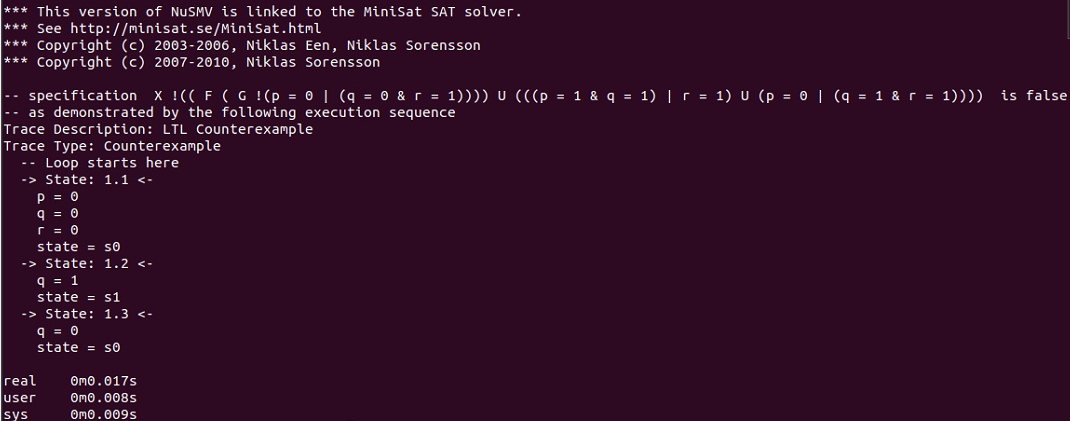}
\caption{another example on MC, \(K_0\) doesn't satisfy the LTL formula \(f_2\), where length of \(f_2\) is 25}
\label{fig4}
\end{figure}

\subsection{Experimental Procedures}

\begin{enumerate}[(1)]
\item 25 LTL formulas f are generated randomly, where the length of each of formula equals to 25. In addition, 25 Kripke structures K are generated randomly. Thus, \(25*25=625\) groups of sub-experiments on NuSMV will be conducted one by one for determining whether or not 25 Kripke structures satisfy 25 LTL formulas, respectively. In fact, we only select 400 groups in all the 625 groups when the length of each formula is 500, as well as 405 groups in all the 625 groups when the length of each formula is 25, due to simplicity.
\item We program on NuSMV for each pair of K and f one by one, and run our program so that the result of model checking is obtained.
\item 405 groups of sub-experiments for model checking produce 405 records, where K and f are filled in the first field and the second one, respectively, in a record, while the model checking result r is filled in the third field in this record. The value of the third field, i.e., r will be 1, if the result of model checking between K and f is true. Otherwise, the value of the third field, i.e., r will be 0. As a result, a data set containing 405 records is obtained. And this is the original data set for our Graph Lab experiments.
\item A part of 405 records will take part in our training process on Graph Lab with BT algorithm. These records form a training set, and other records form a test set. How many records are there in the training set? It depends on the value of some parameters. In fact, we only need adjust the two parameters presented in table 1.
\item We can obtain a ML model M, according to step (4). And we will get the predictive result in terms of model checking if we input the values of the first filed and the second one in a record of the test set, to M.
\item We can compare the predictive result comes from step (5) with the value of the third field, i.e., r. On the basis of it, we can make clear whether the prediction is accurate or not. Furthermore, the average accuracy of prediction can be computed. In this way, we can analysis the power of the new approach.
\item We can obtain and compute the average running time for model checking one pair of K and f on NuSMV, as well as prediction of one pair of K and f on Graph Lab, with the timing function in these two experimental tools. On the basis of it, we can compare the efficiency of the two methods.
From step (1) to step (7), 405 groups sub-experiments on NuSMV and Graph Lab are conducted to study model checking short-length LTL formulas
\item Some LTL formulas f are generated randomly, where the length of each of formula equals to 500. In the similar way with the above procedures, i.e., from step (1) to step (7), we can perform another 400 groups sub-experiments on NuSMV and Graph Lab to study model checking long-length LTL formulas.
\item In the similar way, NuXMV can replace NuSMV to conduct further experiments.
\end{enumerate}

\subsection{Experimental results}
\subsubsection{NuSMV experiments}
{\bf{Example 1}} Figure~\ref{fig2} illustrates an example on Kripke structure \(K=K_0\). \(K_0\) has five states and eight transitions. All three atomic propositions p, q, r are not satisfied in state \(S_0\), while only atomic proposition q is satisfied in state \(S_1\), and so on. The state \(S_0\) can be transformed to state \(S_1\), and state \(S_1\) can be transformed to state \(S_0\) or state \(S_2\), and so on. \(K_0\) can be represented with a string 0000100100101110110122124303243. In this string, the first 15 bits describe whether the three atomic propositions are satisfied in the five states or not, while the rest bits represent the serial number of initial state and the one of final state in the eight transitions.

As for \(K_0=``0000100100101110110122124303243"\) and  \(f_1=``!X((!F((!p\&q|r\\)U(p|!q|r)))U(F(p\&q\&!r))) \) ", the result of NuSMV model checking, i.e., ``true (yes)", is illustrated in Figure~\ref{fig3}, which indicates \(K_0\) satisfy \(f_1\). Therefore, the three fields K, f, r of this record are \(``0000100100101110110122124303243"\), \(``!X((!F((!p\&q|r)U(p|!q|r)))U(F(p\&q\&!r)))"\) and \(``1"\), respectively. Moreover, NuSMV model checking spends 0.018 second this time, as shown in Figure~\ref{fig3}.

As for \(K_0=``0000100100101110110122124303243"\) and  \(f_2=``X!((F(G!(!p|!q\& \\ r)))U((p\&q|r)U(!p|q\&r))) \)", the result of NuSMV model checking, i.e., ``false (no)", is illustrated in Figure~\ref{fig4}, which indicates \(K_0\) does not satisfy \(f_2\). Therefore, the three fields K, f, r of this record are \(``0000100100101110110122124303243"\), \(``X!((F(G!(!p|!q\&r)))U((p\&q|r)U(!p|q\&r)))"\) and \(``0"\), respectively. Moreover, NuSMV model checking spends 0.017 second this time, as shown in Figure~\ref{fig4}.
\begin{table}
  \caption{Graphlab experiments where length of each formula is 25: What will be the values of the parameters if the illustrations of Figure~\ref{fig5-1} occur}
  \label{table1}
  \centering
  \begin{tabular}{lll}
\toprule
    Names of parameters	& Meaning of parameters	& Values of parameters \\
\midrule
   seed&	\makecell[l]{Seed for the random number \\ generator used to split}	& 1988     \\
\midrule
    fraction &	\makecell[l]{For determining the proportion \\ of the records of training set  \\ in the total records of data set}& 0.83 \\
    \bottomrule
  \end{tabular}
\end{table}
Example 1 is over.

Example 1 gives a sample on the two groups of NuSMV model checking experiments. And the two records in the target database are obtained. It should be noted that the database mentioned above is a target database which is the outcome of our model checking experiments. And this database provides raw data for our ML experiments. A data set \(A_1\) containing 405 records is obtained by repeating the similar way with the example 1, where the length of each formula is 25 in this data set. Furthermore, the average running time required for one group of experiment, i.e., model checking for a pair of formula and Kripke structure, is 0.015 second, if the length of each formula is 25.

Similarly, a data set \(A_2\) containing 400 records is obtained by repeating the above way, where the length of each formula is 500 in this data set. Furthermore, the average running time required for one group of experiment, i.e., model checking for a pair of formula and Kripke structure, is 227.28 seconds, if the length of each formula is 500.
\begin{figure}[!htb]
\centering
\subfigure[ experiments on prediction, where length of each formula is 25]{
\begin{minipage}[b]{0.9\textwidth}
\includegraphics[width=\textwidth]{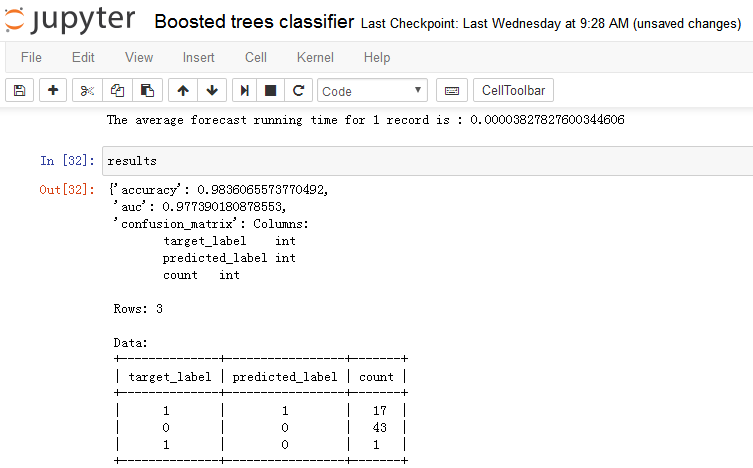} \\
\label{fig5-1}
\end{minipage}
}

\subfigure[experiments on prediction, where length of each formula is 500 ]
{
\begin{minipage}[b]{0.8\textwidth}
\includegraphics[width=\textwidth]{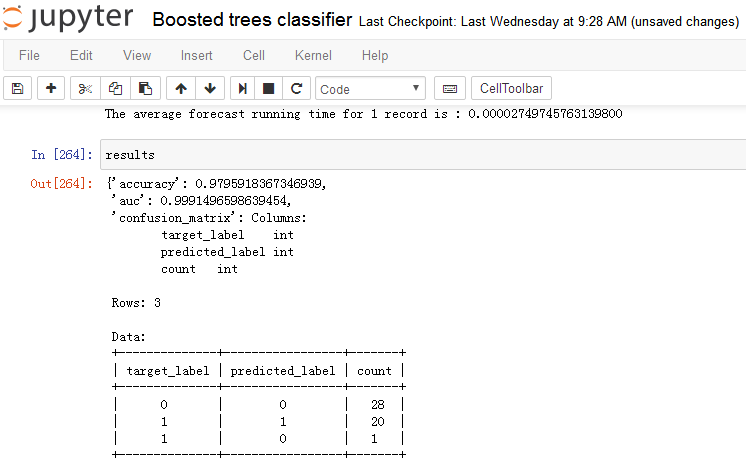} \\
\label{fig5-2}
\end{minipage}
}

\caption{the different predictive results under circumstance of the different lengths of formulas}
\label{fig5}
\end{figure}

\subsubsection{Experiments via Graph Lab}

\begin{table}
  \caption{Graphlab experiments where length of each formula is 500: What will be the values of the parameters if the illustrations of Figure~\ref{fig5-2} occur}
  \label{table2}
  \centering
  \begin{tabular}{lll}
\toprule
    Names of parameters	& Meaning of parameters	& Values of parameters \\
\midrule
   seed&	\makecell[l]{Seed for the random number \\ generator used to split}	& 926     \\
\midrule
    fraction &	\makecell[l]{For determining the proportion \\ of the records  of training set \\ in the total records of data set}	& 0.87 \\
    \bottomrule
  \end{tabular}
\end{table}

The data sets \(A_1\) in section 4.4.1 provides initial data for Grap Lab experiments in section 4.4.2, where the length of each formula is 25. The training set consists of a part of records in data set \(A_1\). And the number of records in the training set can be modified by adjusting some parameters. A ML model M1 can be obtained by employing the training set of \(A_1\) with the BT algorithm. The rest records in \(A_1\) form a test set. For each record, we put the f and K into M1. Comparing the results classified by M1 with r in this record, we can determine whether the predictive result is accurate or not.

\begin{figure}
\centering
\subfigure[predictive accuracy]{
\begin{minipage}[b]{0.4\textwidth}
\includegraphics[width=\textwidth]{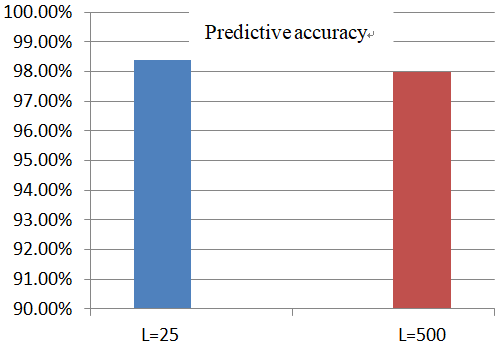} \\
\label{fig6-1}
\end{minipage}
}
\subfigure[average predictive time for one record]
{
\begin{minipage}[b]{0.4\textwidth}
\includegraphics[width=\textwidth]{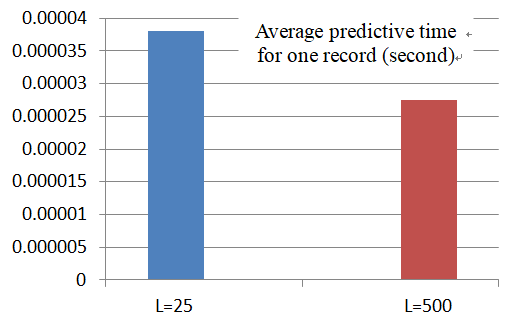} \\
\label{fig6-2}
\end{minipage}
}

\caption{power and efficiency of new method under circumstance of  different lengths of formulas}
\label{fig6}
\end{figure}
Figure~\ref{fig5-1} depicts our optimum result which can be obtained by adjusting the parameters. And the values of the parameters are shown in the Table~\ref{table1}. In this experiment, the accurate rate of the BT algorithm for predicting the results of LTL model checking is 98.4\% and the average time for prediction is 0.000038 second, as shown in the Figure~\ref{fig5-1}. In comparison, the average time consumed by one of 405 groups of NuSMV experiments is 0.015 second. Thus, the efficiency of BT algorithm is \(0.015/0.000038 \approx 395\) times higher than that of the LTL model checking algorithm, if the length of each formula is 25.

The data sets \(A_2\) in section 4.4.1 provides initial data for Grap Lab experiments in section 4.4.2, where the length of each formula is 500. The training set consists of a part of records in data set \(A_2\). And the number of records in the training set can be modified by adjusting some parameters. A ML model M2 can be obtained by employing the training set of \(A_2\) with the BT algorithm. The rest records in \(A_2\) form a test set. For each record, we put the f and K into M2. Comparing the results classified by M2 with r in this record, we can determine whether the predictive result is accurate or not.

Figure~\ref{fig5-2} depicts our optimum result which can be obtained by adjusting the parameters. And the values of the parameters are shown in the Table~\ref{table2}. In this experiment, the accurate rate of the BT algorithm for predicting the results of LTL model checking is 98.0\% and the average time for prediction is 0.0000275 second, as shown in the Figure~\ref{fig5-2}. In comparison, the average time consumed by one of 400 groups of NuSMV experiments is 227.28 seconds. Thus, the efficiency of BT algorithm is \(227.28/0.0000275=8264727\) times higher than that of the LTL model checking algorithm, if the length of each formula is 500.

According to the above results, we can generate a histogram, as shown in Figure~\ref{fig6}.
\begin{table}
  \caption{compared with NuSMV, the new method enhance the efficiency}
  \label{table3}
  \centering
  \begin{tabular}{lllll}
\toprule
   \makecell[l]{Length of \\formulas,\\ i.e., L}	& \makecell[l]{Average running \\time \((t_1)\) of LTL \\model checking for\\ one pair of  Kripke \\ structure and formula}  	& \makecell[l]{Average predictive \\ time \((t_2)\) of \\the  new method  \\ based on BT \\for one record} & \(t_2 / t_1\) & \(t_1 / t_2\) \\
\midrule
  L=25 &	0.015s&	0.000038s&	0.25\%	&395    \\
\midrule
   L=500	& 227.28s&	0.0000275s&	0.000012\%	&8264727 \\
    \bottomrule
  \end{tabular}
\end{table}

\begin{table}
	\caption{compared with NuXMV, the new method enhance the efficiency}
	\label{table4}
	\centering
	\begin{tabular}{lllll}
		\toprule
		\makecell[l]{Length of \\formulas,\\ i.e., L}	& \makecell[l]{Average running \\time \((t_1)\) of LTL \\model checking for\\ one pair of  Kripke \\ structure and formula}  	& \makecell[l]{Average predictive \\ time \((t_2)\) of \\the  new method  \\ based on BT \\for one record} & \(t_2 / t_1\) & \(t_1 / t_2\) \\
		\midrule
		L=25 &	0.017s&	0.000038s&	0.22\%	&447     \\
		\midrule
		L=500	& 258.7s&	0.0000275s&	0.00001063\%	&9407273 \\
		\bottomrule
	\end{tabular}
\end{table}

\subsubsection{NuXMV and Graph Lab experiments}
Now, NuXMV instead of NuSMV is employed by our experiments.

\begin{figure}
	\centering
	\includegraphics[width=\textwidth]{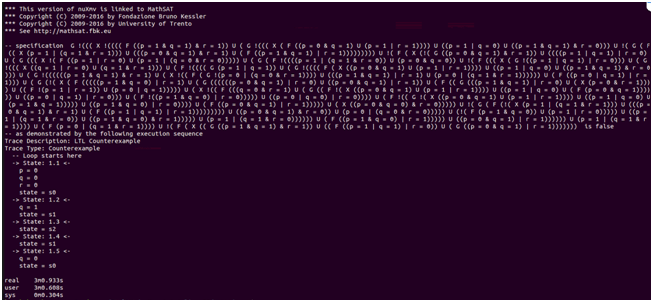}
	\caption{an example on model checking, \(K_0\) doesn’t satisfy the LTL formula \(f_3\), where length of \(f_3\) is 500}
	\label{fig7}
\end{figure}

Figure~\ref{fig7} gives an example on the result of model checking between the Kripke structure \(K_0\) and a LTL formula \(f_3\), using NuXMV. As shown in the figure, the model checking result is false, and the running time is 180.933 seconds.

In our experiments, the average running time via NuXMV for 625 groups of model checking is 0.017 seconds, if the length of each of LTL formulas is 25. And the average running time via NuXMV for 500 groups of model checking is 258.7 seconds, if the length of each of LTL formulas is 500.

Figure~\ref{fig5} gives the running time via the new method. In other words, the average running time of NuXMV-based LTL model checking algorithm is \(0.017/0.000038 \approx 447\) times as much as the new method, if the length of each of LTL formulas is 25, whereas the average running time of NuXMV-based LTL model checking algorithm is 258.7/0.0000275=9407273 times as much as the new method, if the length of each of LTL formulas is 500.

\subsection{Discussion}

First, 130 groups of results are "yes", and other 275 groups of results are "no", for 405 groups of model checking experiments in which the length of each LTL formula is 25. And 141 groups of results are "yes", and other 259 groups of results are "no", for 400 groups of model checking experiments in which the length of each LTL formula is 500. These results provide abundant positive examples and negative ones, which can guarantee the generalization ability of ML model.

Second, as shown in the Figure~\ref{fig6-1}, the predictive accurate of BT algorithm is 98.4\% if the length of the formula is 25, and the predictive accurate of this algorithm is also 98.0\% if the length of the formula is 500. It indicates that the predictions are accurate using the new method based on BT to simulate LTL model checking, regardless of the length of the LTL formula. The LTL model checking is a strongly learnable problem. Therefore, the new method based on BT has a good learning result.

As shown in the Figure~\ref{fig6-2}, the average running time of BT algorithm for predicting one record is less than 0.00005 seconds no matter the length of the formula is 25 or 500. It indicates that the predictions are fast using the new method based on BT to simulate LTL model checking, regardless of the length of the LTL formula. The LTL model checking is a strongly learnable problem, so that it can be simulated in polynomial times.  By contrast, the average running time of LTL model checking algorithm is 258.7 seconds, which is more than 9.4 million times as much as the new method. The reason is that the LTL model checking algorithm has an exponential complexity, while the new method based on BT algorithm has a polynomial complexity.

Third, as shown in Table~\ref{table4}, compared with the LTL model checking algorithm, the new method will enhance the efficiency 447 times if the length of all formulas is 25, whereas the new method will enhance the efficiency 9.4 million times if the length of all formulas is 500. This phenomenon prompts us that the longer the length of the formula, the higher the efficiency of the new method is, due to the advantage of the polynomial algorithm over the exponential algorithm.

Final, as shown in the Figure~\ref{fig6-1}, the accurate rates of the new method for predicting the results of LTL model checking are more than 98.0\%. This phenomenon suggests that the cost of using the new method is acceptable, compared to the significant benefits of using the new method, since the advantage in using the new method is that the max analyzing efficiency is improved more than 9.4 million times, whereas the disadvantage in using the new method is that the analyzing accuracy will go down by only 2 percent, if the length of the formula is 500.

\section{Conclusions}
In this paper, the machine learning technique is introduced to predict results of model checking.  On the basis of it, an approximate model checking method is formed. To the best of our knowledge, this is the first ML-based approximate model checking approach. The core of the existing model checking methods is to explore exhaustively all the states. By contrast, our new method based on BT algorithm does not search state space. As a result, the state explosion problem and the exponential complexity are avoided. Furthermore, the new method based on BT complements the existing ones. The new method has an acceptable accuracy of approximate model checking which has declined slightly from that of actual model checking, in exchange for a substantial increase in efficiency of model checking. As a result, the longer the LTL formula, the more obvious the comparative advantage of the new methods is. This is the benefit of using the new method.

\section*{Acknowledgements}
This work has been supported by the National Natural Science Foundation of China (No.U1204608).

\section*{References}

\medskip

\small
[1] Clarke E, et a1. Model Checking. Massachusetts: MITPress, 1999.

[2] Barnat J, Bauch P, Brim L, et a1. Designing fast LTL model checking algorithms for many-core GPUs. {\it Journal of Paralleland Distributed Computing}, 2012, {\bf 72}(9): 1083-1097.

[3] Carbone R. LTL model-checking for security protocols. {\it AI Communications}, 2011, {\bf 24}(4): 281-283.

[4] C Song F, Touili T. Model-Checking for Android Malware Detection, {\it Programming Languages and Systems}. Springer International Publishing, 2014:216-235.

[5] Clarke E, Grand challenge problem: Model check concurrent software. \\ http://www.csl.sri.com/users/shankar/VGC05/Clarke.ppt

[6] Pnueli A. The temporal logic of programs. {\it Proceedings of the 18th Annual Symposium on Foundations of Computer Science}. Washington, USA, 1977: 46-57.

[7] Benari M, Pnueli A, Manna Z. The temporal logic of branching time. {\it Acta Informatica}, 1983, {\bf 20}(3): 207-226.

[8] Emerson E, Clarke E. Using branching time temporal logic to synthesize synchronization skeletons. {\it Science of Computer Programming}, 1982, {\bf 2}(3): 241-266.

[9] Burch J R, Clarke E M, Long D E, et al. Symbolic Model Checking for Sequential Circuit Verification. {\it IEEE Trans. Comput. -Aided Des}. Integrated Circuits \& Syst. 1994, {\bf 13(}4):401-424.

[10] http://nusmv.fbk.eu/

[11] Peng Xujun, Setlur Srirangaraj, Govindaraju Venu, Using a boosted tree classifier for text segmentation in hand-annotated documents, {\it Pattern Recognition Letters}, v 33, n 7, p 943-950, May 1, 2012.

[12] Demirkir Cem, Sankur Bülent, Face detection using boosted tree classifier stages, {\it Proceedings of the IEEE 12th Signal Processing and Communications Applications Conference}, SIU 2004, p 575-578, 2004.

[13] Parag Toufiq, Elgammal Ahmed, Unsupervised learning of boosted tree classifier using graph cuts for hand pose recognition, {\it Proceedings of the British Machine Vision Conference 2006}, p 1259-1268, 2006, BMVC 2006.

[14] Wu Bo, Nevatia Ram, Cluster boosted tree classifier for multi-view multi-pose object detection, {\it Proceedings of the IEEE International Conference on Computer Vision, 2007}.

[15] Day Matthew, Emotion recognition with boosted tree classifiers, {\it Proceedings of the 2013 ACM International Conference on Multimodal Interaction, p 531-534, 2013}.

[16] Friedman JH. Greedy function approximation: a gradient boosting machine. {\it Annal of Statistics}, 2001, (29): 1189–1232

[17] http://scikit-learn.org/stable/modules/ensemble.html\#gradient-boosting

[18] https://turi.com/

\end{document}